# Integrated nanophotonic polarizers in silicon waveguides and ring resonators using graphene oxide 2D films

David J. Moss[1, a)]

[1] Optical Sciences Centre, Swinburne University of Technology, Hawthorn, Victoria 3122, Australia

[a)] dmoss@swin.edu.au

**Abstract**

We experimentally demonstrate waveguide and microring resonator (MRR) polarizers by integrating 2D graphene oxide (GO) films onto silicon (Si) photonic devices. The 2D GO films with highly anisotropic light absorption are on-chip integrated with precise control over their thicknesses and sizes. Detailed measurements are performed for the fabricated devices with different GO film thicknesses, coating lengths, and Si waveguide widths. The results show that a maximum polarization-dependent loss (*PDL*) of ~17 dB is achieved for the hybrid waveguides, and the hybrid MRRs achieved a maximum polarization extinction ratio (*PER*) of ~10 dB. We also characterize the wavelength- and power-dependent response for these polarizers. The former demonstrates a broad operation bandwidth of over ~100 nm, and the latter verifies performance improvement enabled by photo-thermal changes in GO films. By fitting the experimental results with theoretical simulations, we find that the anisotropy in the loss of GO films dominates the polarization selectivity of these devices. These results highlight the strong potential of 2D GO films for realizing high-performance polarization selective devices in Si photonic platform.

Optical polarizers transmit light only polarised in one direction while blocking perpendicularly polarized light. They are important functions for modern optical systems.[1, 2] They have facilitated polarization selection and control in a wide range of applications, such as optical sensing,[3, 4] imaging and display,[5, 6] optical analysis,[7, 8] and optical communications.[9, 10]

A variety of optical polarizers have been implemented based on different bulk material device platforms, such as refractive prisms,[11, 12] birefringent crystals,[13, 14] fiber components,[15, 16] and integrated photonic devices.[17, 18] Amongst these, integrated photonic polarizers, particularly those based on the well-developed silicon (Si) photonic platform, offer attractive advantages of compact footprint, low power consumption, and the capability for large-scale manufacturing. Nevertheless, the rapid advancement of photonics industries drives the demand for high-performance optical polarizers across broad wavelength ranges, which is usually challenging for optical polarizers based on bulk materials.[19-21]

Recently, owing to their high anisotropy in light absorption and broadband response, 2D materials, such as graphene,[22, 23] graphene oxide (GO),[24, 25] transition-metal dichalcogenides (TMDCs),[26, 27] and black phosphorus,[28] have been on-chip integrated to realize optical polarizers with high performance and new features. As a common derivative of graphene, GO with facile fabrication processes for precise on-chip integration shows a high compatibility with integrated device platforms.[29-31] Previously, we reported high performance waveguide and microring resonator (MRR) polarizers by integrating 2D GO films onto high index doped silica devices.[25] In this work, we further integrate 2D GO films onto the more widely used Si photonic devices to realize waveguide and MRR polarizers. Precise control of the GO film thicknesses is achieved by using a transfer-free and layer-by-layer film coating method, together with window opening on the upper cladding of Si photonic devices to control the film coating lengths. We perform detailed measurements for the fabricated devices with different GO film thicknesses, coating lengths, and Si waveguide widths, achieving a maximum polarization-dependent loss (*PDL*) of ~17 dB for the waveguide polarizers and a maximum polarization extinction ratio (*PER*) for the MRR polarizers. The wavelength- and power-dependent response of these polarizers is also characterized, showing a broad operation bandwidth over ~100 nm and improvement in the polarization selectivity enabled by photo-thermal changes in GO films. Finally, we perform theoretical analysis by fitting the experimental results with theory, which reveals that the anisotropy in the loss of GO films dominates the polarization selectivity of these polarizers. These results verify the effectiveness of integrating 2D GO films onto Si photonic devices for implementing high-performance optical polarizers.

**Figure 1(a)** illustrates the atomic structure of monolayer GO, in which the carbon network is decorated with diverse oxygen functional groups (OFGs) such as epoxide, hydroxyl, and carboxylic groups.[32, 33] Unlike graphene that has a zero bandgap and exhibits metallic behavior,[34] GO is a dielectric material with an opened bandgap typically ranging between ~2.1 and ~3.6 eV.[30, 32] This bandgap is larger than the energy of two photons at 1550 nm (*i.e.*, ~1.6 eV), which yields both low linear light absorption and two-photon absorption (TPA) in the telecom band.

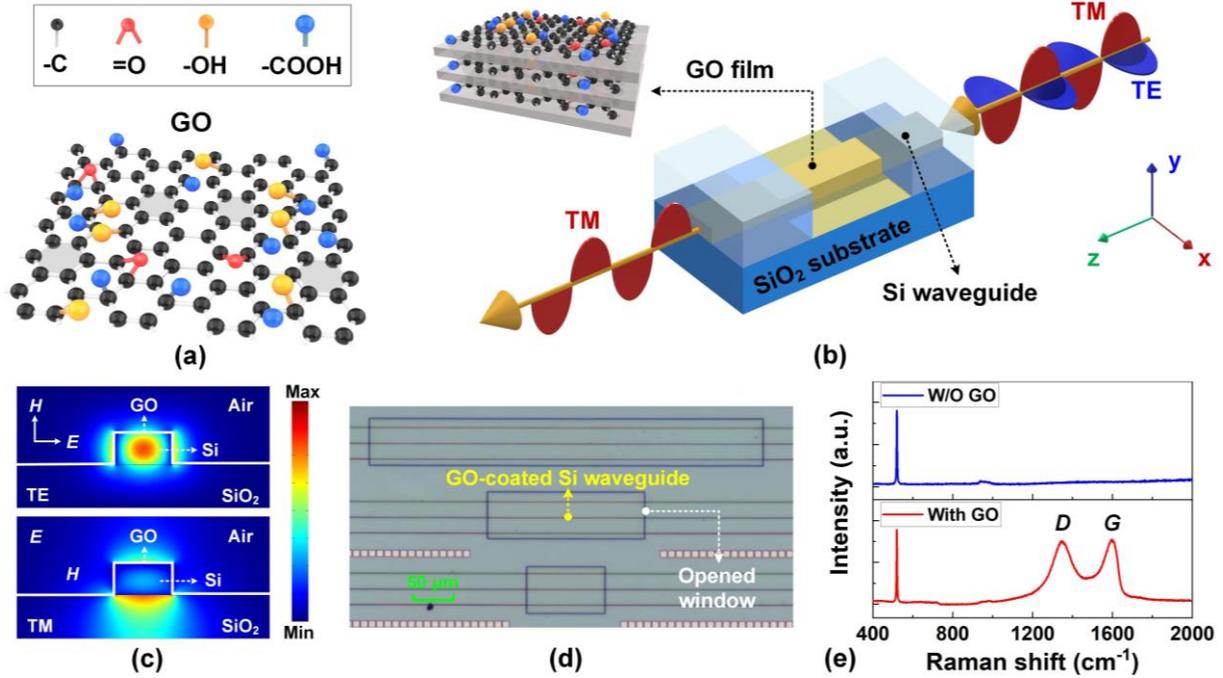

FIG. 1. (a) Schematic of GO's atomic structure. (b) Schematic illustration of a waveguide optical polarizer based on a Si waveguide integrated with a monolayer GO film. Inset illustrates the layered structure of GO films fabricated by the self-assembly method. (c) TE and TM mode profiles for the hybrid waveguide in (b). (d) Microscopic image of the fabricated devices on a silicon-on-insulator (SOI) chip coated with a monolayer GO film. (e) Measured Raman spectra of the SOI chip in (d) before and after coating the GO film.

**Figure 1(b)** shows the schematic of a Si nanowire waveguide integrated with a monolayer GO film. The corresponding transverse electric (TE) and transverse magnetic (TM) mode profiles for the hybrid waveguide are shown in **Fig. 1(c)**, which were simulated using a commercial mode solving software (COMSOL Multiphysics). The width and height of the Si waveguide are 400 nm and 220 nm, respectively. Due to the interaction between the evanescent field from the Si waveguide and the 2D GO film with strong anisotropy in its light absorption, the hybrid waveguide exhibits much stronger light absorption for TE (in-plane) polarization than TM (out-of-plane) polarization. This enables it to function as a TM-pass optical polarizer.

**Figure 1(d)** shows a microscopic image of the fabricated devices on a silicon-on-insulator (SOI) chip. The device layout was patterned via 248-nm deep ultraviolet lithography followed by inductively coupled plasma etching. A silica upper cladding was deposited onto the fabricated Si waveguides. To enable light-GO interaction, lithography and dry etching processes were employed to open windows on the silica upper cladding to allow the coating of GO films onto the Si waveguides. We fabricated devices with 5 different lengths (ranging between ~0.1 mm and ~2.2 mm) for the opened windows, which correspond to different GO film coating lengths in the hybrid waveguides. The GO films were coated layer-by-layer and transfer-free with a solution-based self-assembly process.[25, 31] This coating process has significant advantages over commonly used mechanical transfer methods for coating 2D materials,[35, 36] achieving large-area conformal coating onto nanostructures, precise film thickness control, and high film uniformity.[29, 30] As shown in **Fig. 1(d)**, the good morphology of the coated GO films provides evidence for the high film uniformity achieved by our coating method.

**Figure 1(e)** shows the measured Raman spectra of the SOI chip in **Fig. 1(d)** before and after coating a monolayer GO film. The GO film had a thickness of ~2 nm, which was characterized by atomic force microscopy measurement. In the Raman spectrum for the GO-coated chip, the presence of the representative $D$ (1345 cm$^{-1}$) and $G$ (1590 cm$^{-1}$) peaks verifies the successful on-chip integration of the GO film.

In **Fig. 2**, we show the results for the measured insertion losses ($IL$'s) of the fabricated devices for input light with different polarization states. We measured devices with different GO layer numbers ($N$), GO coating lengths ($L_{GO}$), and Si waveguide widths ($W$). For all the devices, the total length of the Si waveguides was ~3.0 mm. In our measurements, a continuous-wave (CW) light at ~1550 nm was butt coupled into / out of the devices via lensed fibers. The fiber-to-chip coupling loss was ~5 dB per facet. For comparison, the input power was kept the same as $P_{in}$ = ~0 dBm. Unless otherwise specified, the values of $P_{in}$ and $IL$ in our discussion refer to those after excluding the fiber-to-chip coupling loss.

**Figures 2(a-i)** and **2(a-ii)** show the measured *IL* versus GO coating length $L_{GO}$ for TE- and TM-polarized input light, respectively. Here we show the results for the hybrid waveguides with 1 – 5 layers of GO (*i.e.*, $N$ = 1 – 5). For comparison, all the uncoated Si waveguides had the same waveguide width of $W$ = ~400 nm. In our measurements, three duplicate devices were measured, and the data points depict their average values, with the error bars illustrating the variations for different samples. It can be seen that the *IL* increases with $L_{GO}$ and $N$ for both polarizations, with the TE polarization showing a more significant increase than the TM polarization. In **Fig. 2(a-iii)**, we further calculated the *PDL* (dB) by subtracting the TM-polarized *IL* (dB) from the TE-polarized *IL* (dB). As can be seen, the *PDL* increases with $L_{GO}$ and $N$. For the 5-layer device with $L_{GO}$ = ~2.2 mm, a maximum *PDL* value of ~17 dB was achieved. In contrast, the uncoated Si waveguide did not show any significant polarization dependent *IL*, with the *PDL* being less than 0.2 dB. The difference in the TM-polarized *IL* values between the uncoated and hybrid waveguides reflects the minimum excess insertion loss induced by the GO film. For the 5-layer device with $L_{GO}$ = ~2.2 mm, the difference was ~10 dB.

**Figures 2(b-i)** and **2(b-ii)** show the measured *IL* versus $W$ for TE- and TM-polarized input light, respectively. Here, all the hybrid waveguides had the same $L_{GO}$ = ~2.2 mm. For both polarizations, the *IL* increases with GO layer number $N$. In contrast to the results in **Figs. 2(a-i)** and **2(a-ii)**, the *IL* decreases as $W$ increases, with the TE polarization showing a more significant decrease than the TM polarization. This results in the decrease of *PDL* for an increasing $W$ in **Fig. 2(b-iii)**. For the 5-layer device with $W$ = ~600 nm, the *PDL* was ~5 dB, which was ~12 dB lower than that of the comparable device with $W$ = ~400 nm.

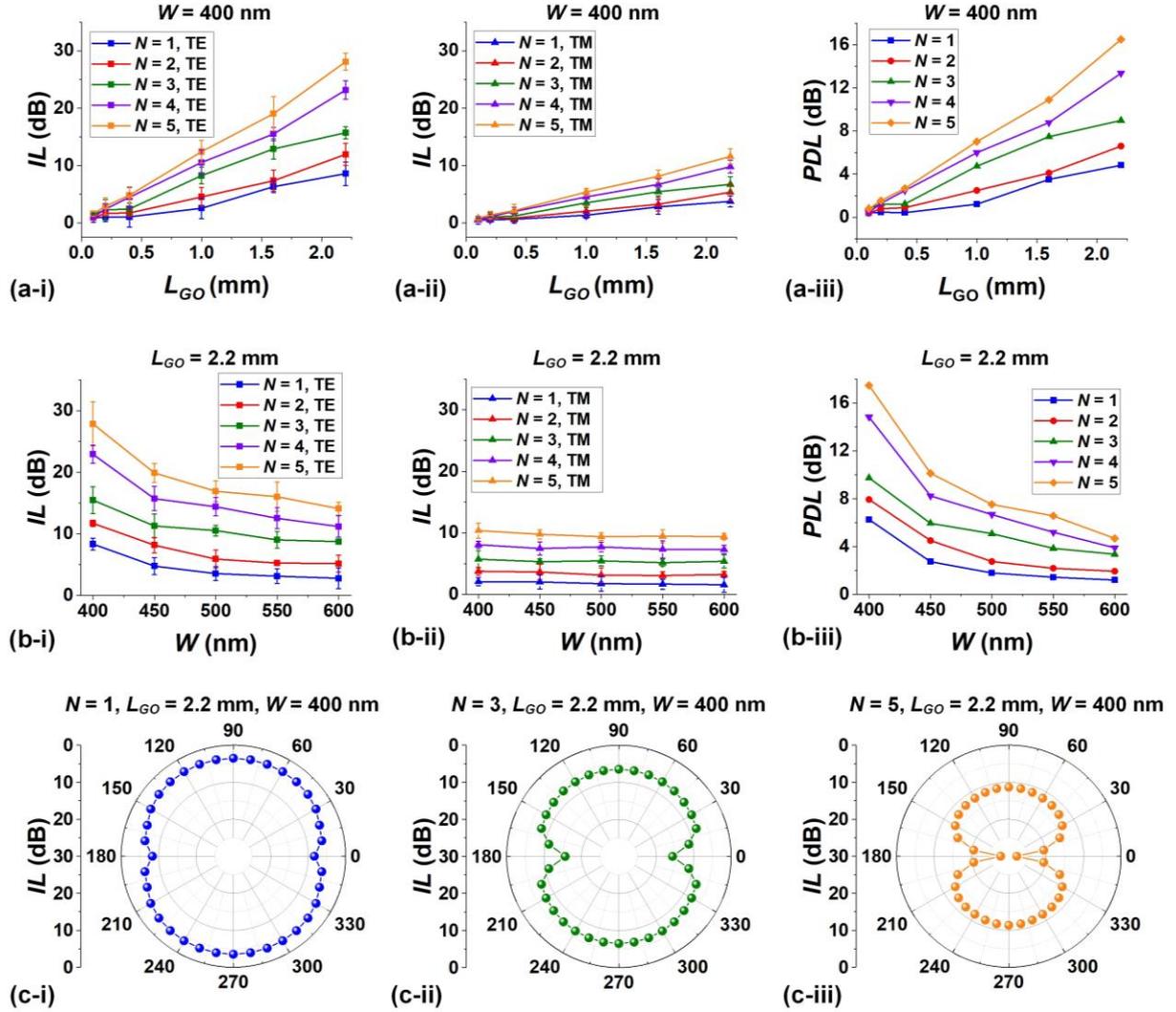

FIG. 2. (a) Measured (i) TE- and (ii) TM-polarized insertion loss (*IL*) versus GO coating length $L_{GO}$ for the hybrid waveguides with 1−5 layers of GO. (iii) shows the polarization dependent loss (*PDL*) calculated from (i) and (iii). (b) Measured (i) TE-, (ii) TM- polarized *IL* versus Si waveguide width *W* for the hybrid waveguides with 1−5 layers of GO. (iii) shows the *PDL* calculated from (i) and (iii). (c) Polar diagrams for the measured *IL* of devices with different GO layer numbers of (i) $N = 1$, (ii) $N = 3$, and (iii) $N = 5$. The polar angle represents the angle between the input polarization plane and the substrate. In (a) – (c), the input CW power and wavelength were ~0 dBm and ~1550 nm, respectively. In (a) and (b), the data points depict the average of measurements on three duplicate devices and the error bars illustrate the variations among the different devices. In (a), $W$ = ~400 nm. In (b), $L_{GO}$ = ~2.2 mm. In (c), $L_{GO}$ = ~2.2 mm and $W$ = ~400 nm.

**Figure 2(c)** shows the polar diagrams for the measured *IL* of devices with different GO layer numbers (*i.e.*, $N = 1$, 3, and 5). Except for the GO layer number, the other device parameters were kept the same (*i.e.*, $W$ = ~400 nm and $L_{GO}$ = ~2.2 mm). In the polar diagrams, the variations in the *IL* values across various polarization angles further confirm the polarization selection capability for the hybrid waveguides. Compared to the 5-layer device that achieved a *PDL* of ~17 dB, the 1-layer and 3-layer devices achieved lower *PDL* values of ~5 dB and ~9 dB, respectively.

**Figure 3a** shows the measured *PDL* versus wavelength of the input CW light. In **Fig. 3(a-i)**, we show the results for the hybrid waveguides with 1 − 5 layers of GO (*i.e.*, $N$ = 1 − 5) and all of them had the same $L_{GO}$ = ~2.2 mm and $W$ = ~400 nm. In **Fig. 3(a-ii)**, we show the results for the hybrid waveguides with different $L_{GO}$ but the same $N$ = 5 and $W$ = ~400 nm. For all the devices in **Fig. 3a**, the *PDL* shows a very small variation that was less than 1 dB within the measured wavelength range of ~1500 − 1600 nm. This reflects the broad operation bandwidth for these waveguide polarizers. We note that there was a slight increase in the *PDL* as the wavelength increased, which can be attributed to a minor change in GO's mode overlap induced by dispersion. In our measurements, the wavelength tuning range was limited by the employed tunable CW laser. The light absorption bandwidth for GO is in fact quite broad, which covers infrared wavelengths and extends into the visible and terahertz (THz) ranges.[29, 37] Such broadband response enables a much broader operation bandwidth (potentially spanning several hundreds of nanometers, as we demonstrated in Ref. [25]) than that demonstrated here. This represents a significant advantage of 2D-material-based polarizers, which is usually challenging to achieve for integrated photonic devices based on only bulk materials.[2, 38]

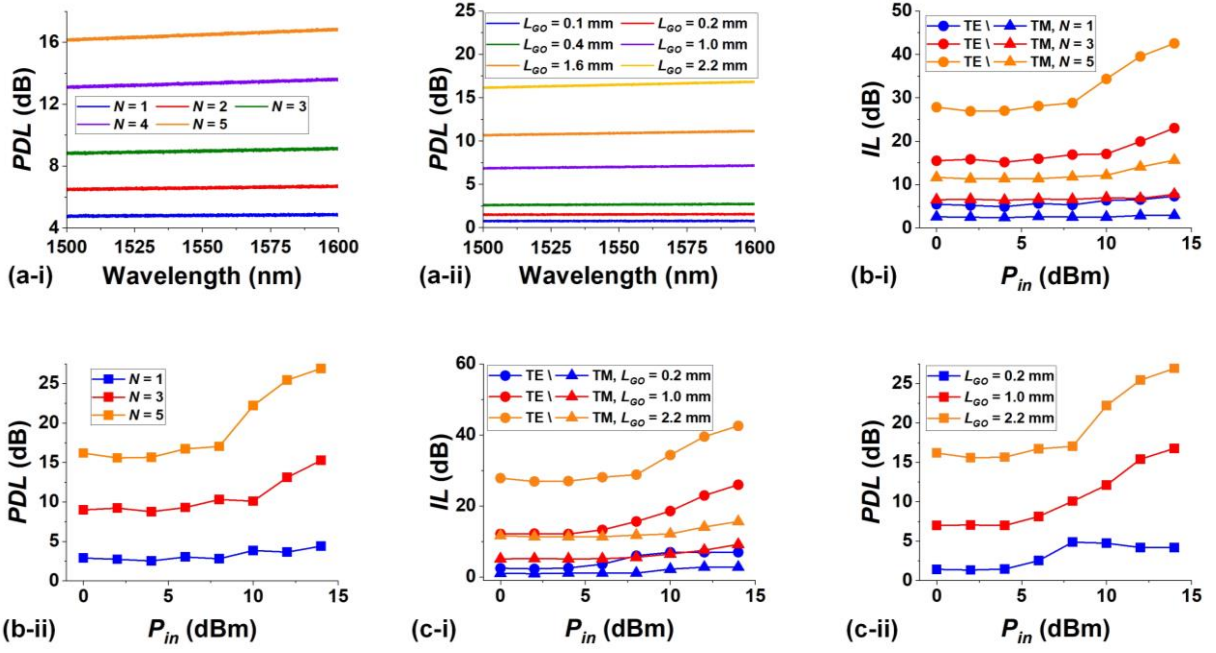

FIG. 3. (a) Measured *PDL* versus wavelength of input CW light for (i) devices with different GO layer numbers ($N$) but the same GO coating length ($L_{GO}$) of ~2.2 mm and (ii) devices with different $L_{GO}$ but the same $N$ = 5. (b) (i) Measured TE and TM polarized *IL* versus input power ($P_{in}$) for devices with different $N$ = 1, 3, and 5 but the same $L_{GO}$ = ~2.2 mm. (ii) shows the *PDL* calculated from (i). (c) (i) Measured TE and TM polarized *IL* versus $P_{in}$ for devices with different $L_{GO}$ = 0.2, 1.0, and 2.2 mm but the same $N$ = 5. (ii) shows the *PDL* calculated from (i). In (a) – (c), $W$ = ~400 nm. In (a), $P_{in}$ = ~0 dBm. In (b) and (c), the input CW wavelength was ~1550 nm.

**Figure 3(b-i)** shows the measured *IL* versus input power $P_{in}$ for the hybrid waveguides with 1, 3, and 5 layers of GO. Here all the devices had the same $L_{GO}$ = ~2.2 mm and $W$ = ~400 nm. As $P_{in}$ increases, the *IL* increases for both polarizations, with the TE polarization showing a more significant increase than the TM polarization. The increased *IL* was induced by the photo-thermal reduction of GO at high light powers, where reduced GO exhibited stronger light absorption than unreduced GO.[30, 39] Another interesting feature for the reduction of GO induced by photo-thermal effects is its reversibility within a certain power range.[39, 40] This originates from the instability of photo-thermally reduced GO, which can readily return to the unreduced state after cooling in oxygen containing atmosphere (*e.g.*, after tuning off the optical input). Since the hybrid waveguides exhibited stronger absorption for TE-polarized light, the reduction of GO occurred more readily for TE polarization, resulting in a more significant change in the corresponding *IL* in **Fig. 3(b-i)**. We also note that the 5-layer device showed more significant changes in the *IL* as compared to the 3-layer and 1-layer devices, which reflects the fact that there were more significant photo-thermal effects in thicker GO films. In **Fig. 3(b-ii)**, we show the calculated *PDL* versus $P_{in}$, where the *PDL* increases for an increasing $P_{in}$. This indicates that the polarization selectivity was further improved by increasing the input power. At $P_{in}$ = ~15 dBm, the *PDL* for the 5-layer device was ~25 dB, representing an ~8-dB improvement relative to the *PDL* at $P_{in}$ = ~0 dBm.

**Figure 3(c-i)** shows the measured *IL* versus $P_{in}$ for the hybrid waveguides with three different $L_{GO}$ but the same $N$ = 5 and $W$ = ~400 nm. The extracted *PDL* versus $P_{in}$ is shown in **Fig. 3(c-ii)**. For all the devices, the *PDL* starts to increase at almost the same power threshold of $P_{in}$ = ~6 dB. After that, the devices with $L_{GO}$ = ~1.0 mm and ~2.2 mm exhibit a similar rate of increase, whereas the rate for the device with $L_{GO}$ = ~0.2 mm gradually decreases. This reflects an interesting fact that the reduction of GO induced by the photo-thermal effects was non-uniform along the waveguide. It first occurred at the start of the GO section near the input port and became weaker in tandem with the attenuation of the light power along the waveguide.

Based on the results in **Figs. 2** and **3**, we further analyze the properties of 2D GO films by fitting the experimental results with theoretical simulations. **Figure 4(a-i)** shows the propagation losses (*PL*'s) of the hybrid waveguides versus GO layer number *N* for both polarizations, which were extracted from the measured *IL*'s in **Figs. 2(a-i)** and **2(a-ii)**. **Figure 4(a-ii)** shows the extinction coefficients (*k*'s) of 2D GO films obtained by fitting the *PL*'s in **Fig. 4(a-i)** with optical mode simulations, and the ratios of TE- to TM-polarized *k* values are further plotted in **Fig. 4(a-iii)**. For all different *N*, the GO films exhibit much larger *k* values for TE polarization than TM polarization in **Fig. 4(a-ii)**, highlighting the high anisotropy in their light absorption. For both polarizations, there is a slight increase in *k* as *N* increases. This can possibly be attributed to more significant scattering loss in thicker GO films, which can be induced by increased film unevenness and imperfect contact between adjacent layers. It is also interesting to note that in **Fig. 4(a-iii)** the ratio of the *k* values remains relatively consistent without any significant variations.

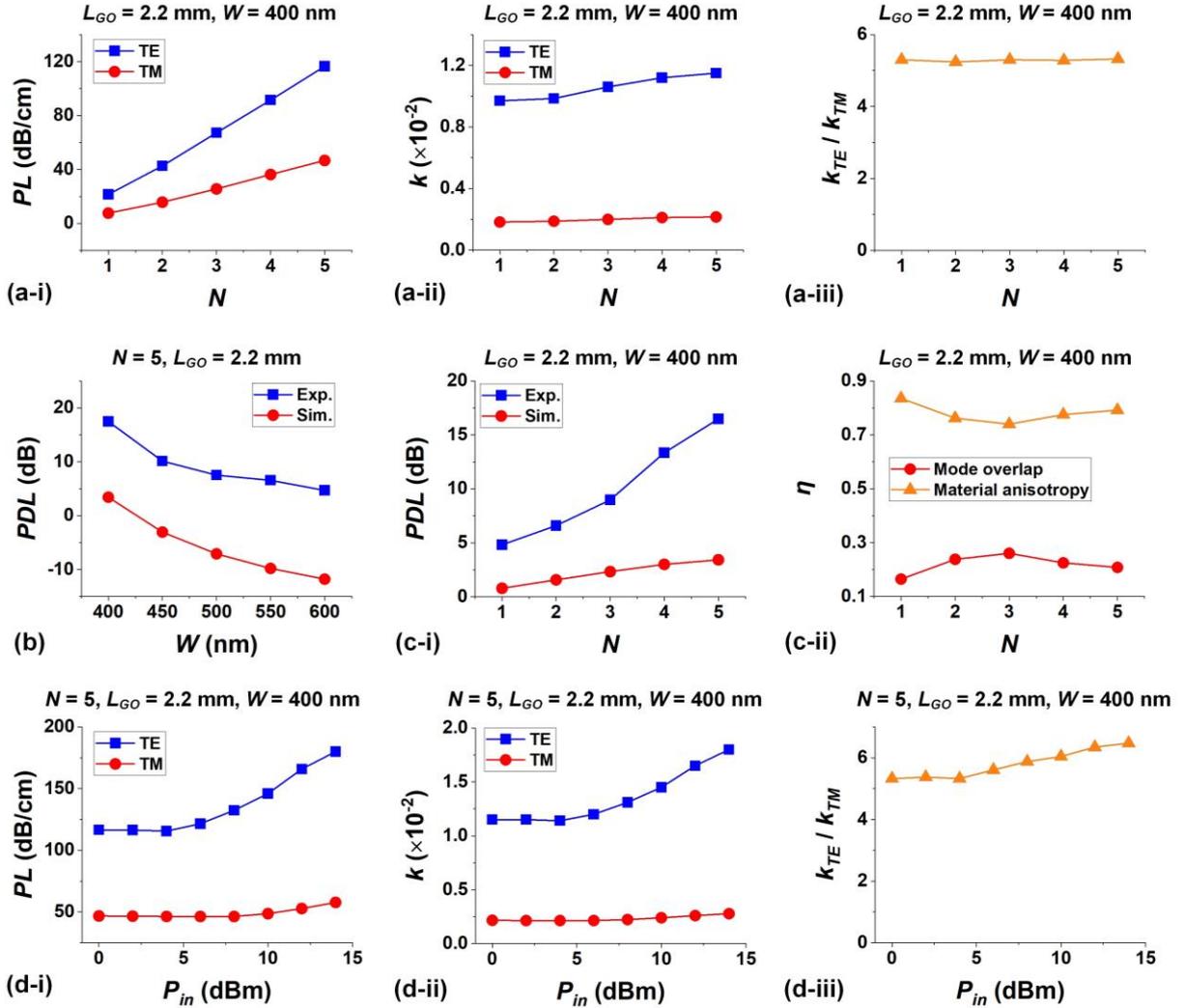

FIG. 4. (a-i) TE- and TM-polarized waveguide propagation loss (*PL*) versus GO layer number (*N*) for the hybrid waveguides with 1−5 layers of GO. (a-ii) Extinction coefficients (*k*'s) of 2D GO films obtained by fitting the results in (a-i) with optical mode simulations. (a-iii) Ratios of *k* values for TE and TM polarizations ($k_{TE} / k_{TM}$) extracted from (a-ii). (b) Measured (Exp.) and simulated (Sim.) *PDL* versus Si waveguide width (*W*) for the hybrid waveguides with 5 layers of GO. The simulated *PDL* values were obtained by using the same *k* value for both TE and TM polarizations. (c-i) Measured (Exp.) and simulated (Sim.) *PDL* versus *N*. (c-ii) Fractional contributions (*η*'s) to the overall *PDL* from polarization-dependent mode overlap and material loss anisotropy, which were extracted from (c-i). (d-i) TE- and TM-polarized *PL* versus input power ($P_{in}$) for the hybrid waveguide with 5 layers of GO. (d-ii) *k*'s of 2D GO films obtained by fitting the results in (d-i) with optical mode simulations. (d-iii) $k_{TE} / k_{TM}$ extracted from (d-ii). In (a) – (d), the GO coating length $L_{GO}$ = ~2.2 mm. In (a) – (c), $P_{in}$ = ~0 dBm. In (a), (c), and (d), *W* = ~400 nm.

In **Fig. 4b**, we compare the measured *PDL* values with those obtained from optical mode simulations. In our simulations, we assumed that the GO films were isotropic with the same *k* value (*i.e.*, $k_{TE}$ for $N = 5$ in **Fig. 4(a-ii)**) for both TE and TM polarizations. As a result, the simulated *PDL* values represent the polarization selectivity enabled by the polarization-dependent GO mode overlap, and the difference between the measured and simulated *PDL* values reflects the extra polarization selectivity provided by the loss anisotropy of GO films. The simulated *PDL* values are negative when $W > 400$ nm. This is because the polarization-dependent GO mode overlap induces a higher loss for TM polarization. Despite this, the net *PDL* values remain positive, indicating that the material loss anisotropy compensates for the negative *PDL* induced by GO mode overlap.

Similar to that in **Fig. 4b**, in **Fig. 4(c-i)** we show the measured and simulated *PDL* values for the devices with different *N* but the same $W = \sim 400$ nm. Unlike that in **Fig. 4b**, both the measured and simulated *PDL*'s show positive values for all different *N*. This suggests that the polarization-dependent GO mode overlap also contributes to the overall *PDL*. In **Fig. 4(c-ii)**, we further calculated the fractional contributions to the overall *PDL* from the polarization-dependent mode overlap and the material loss anisotropy (where the sum of the two fractions equals to 1). As can be seen, the contribution from the material loss anisotropy, which accounts for more than 70% for all different GO layer numbers, dominates the overall *PDL*. This further highlights the significance of the anisotropic 2D GO films in facilitating the functionality of the polarizer.

**Figure 4(d-i)** shows the *PL*'s of the hybrid waveguides versus input CW power $P_{in}$ for both polarizations, which were extracted from the measured *IL*'s in **Fig. 3b**. **Figure 4(d-ii)** shows the *k* values of 2D GO films obtained by fitting the *PL*'s in **Fig. 4(d-i)** with optical mode

simulations, and **Fig. 4(d-iii)** shows the ratios of the TE- to TM-polarized *k* values in **Fig. 4(d-ii)**. In **Fig. 4(d-ii)**, *k* increases with $P_{in}$ for both polarizations, with the TE polarization showing a more significant increase than the TM polarization. In **Fig. 4(d-iii)**, the ratio between the *k* values slightly increases with $P_{in}$, which is also resulting from the more significant photo-thermal effects in GO films induced by stronger absorption of TE-polarized light.

Except for waveguide optical polarizers, we also integrated 2D GO films onto Si MRRs to implement polarization-selective MRRs. **Figure 5(a)** shows the device schematic, and **Fig. 5(b)** shows a microscopic image of the fabricated device with a monolayer GO film. The MRRs were fabricated together with the waveguides in **Fig. 1(d)** via the same processes. The rings and the bus waveguides had the same width of *W* = ~400 nm. The radius of the MRRs was ~20 µm, and the length of the opened windows on the MRRs was ~10 µm.

By scanning the wavelength of a CW light coupled into the bus waveguides, we measured the TE- and TM-polarized transmission spectra of the MRRs. **Figure 5(c)** shows the measured spectra for the hybrid MRRs with 1 and 2 layers of GO, together with those for the uncoated MRR for comparison. For all the measured spectra, the input CW power remained constant at $P_{in}$ = ~-10 dBm. **Figure 5(d)** shows the extinction ratios (*ER*'s) of these MRRs extracted from **Fig. 5(c)**. For the uncoated MRR, the *ER*s for TE and TM polarizations were ~33 dB and ~32 dB, respectively. In contrast, the *ER*s decreased to ~16 dB and ~26 dB for the 1-layer device, and ~11 dB and ~19 dB for the 2-layer device. **Figure 5(e)** shows the polarization extinction ratio (*PER*) calculated from **Fig. 5(d)**, which is defined as the absolute difference between the *ER*'s of the TE- and TM-polarized resonances. As can be seen, the uncoated MRR exhibited negligible polarization selectivity, with its *PER* being less than ~1 dB. In contrast, the hybrid devices with 1 and 2 layers of GO exhibited *PER* values of ~10 dB and ~8 dB, respectively, highlighting their high polarization selectivity. Compared to the device with 2 layers of GO, the device with 1 layer of GO exhibited a higher *PER*. This is mainly induced by a more significant decrease in the TM-polarized *ER* for the device with a thicker GO film.

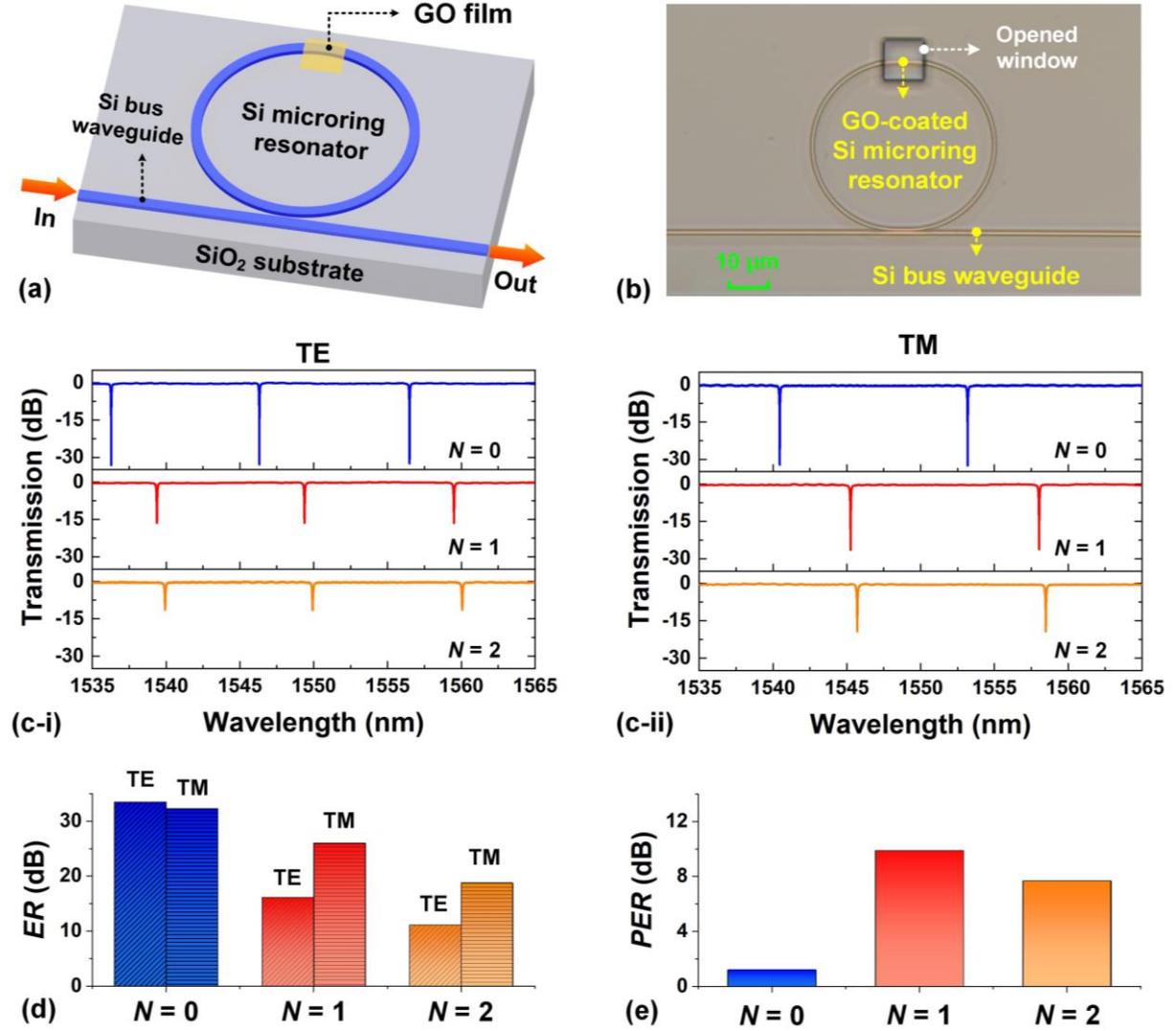

FIG. 5. (a) Schematic illustration of a GO-coated Si microring resonator (MRR) as a polarization-selective MRR. (b) Microscopic image of a fabricated Si MRR coated with a monolayer GO film. (c) Measured (i) TE- and (ii) TM-polarized transmission spectra of the hybrid MRRs with 1 and 2 layers of GO ($N$ = 1, 2). The corresponding results for the uncoated MRR ($N$ = 0) are also shown for comparison. (d) Extinction ratios (*ER*'s) for the MRRs extracted from (c). (e) Polarization extinction ratios (*PER*'s) extracted from (d). In (c) – (e), the CW input power was $P_{in}$ = ~-10 dBm.

In **Fig. 6**, we characterize the power-dependent response for the polarization-selective MRR. We employed two CW inputs in our measurements. On CW beam provided a pump into one of the MRR's tuned to a resonance. Its wavelength was adjusted near the resonance until a thermal equilibrium state was reached. Next, a second lower CW power beam at -10 dBm was used to probe the MRR's transmission spectra. In contrast to directly using a high-power CW pump to scan the spectra, the probe minimized any resonance lineshape asymmetry due to optical bistability,[41, 42] thus enabling more precise characterization of the *ER*s.

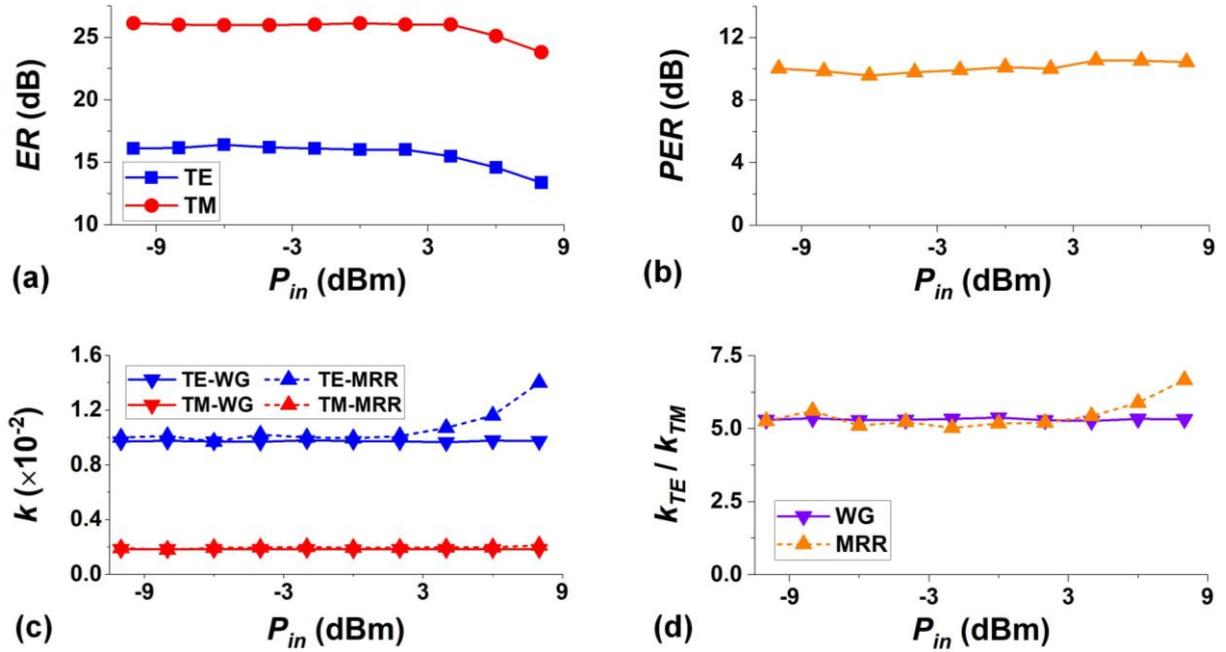

FIG. 6. (a) TE- and TM-polarized *ER* versus input power ($P_{in}$) for the hybrid MRR with 1 layer of GO ($N = 1$). (b) *PER* extracted from (a). (c) Fit $k$'s obtained from the MRR experiment and the waveguide (WG) experiment. (d) Ratios of $k$ values for TE and TM polarizations ($k_{TE} / k_{TM}$) extracted from (c).

In **Fig. 6(a)**, we plot TE- and TM-polarized *ER*'s versus input pump power $P_{in}$ for the hybrid MRR with 1 layer of GO. **Figure 6(b)** shows the *PER* calculated from **Fig. 6(a)**. For TE polarization in **Fig. 6(a)**, the *ER* remains constant at ~17 dB when $P_{in} \leq$ ~0 dBm. When $P_{in} >$ ~0 dBm, there is an obvious decrease in the *ER* resulting from an increased loss induced by photo-thermal reduction of GO. For TM polarization, the *ER* remains constant at ~26 dB when $P_{in} \leq 4$ dBm, followed by a decrease when $P_{in} > 4$ dBm. Compared to TE polarization, TM polarization exhibits a higher power threshold at which the *ER* starts to decrease. This further confirms the relatively weak photo-thermal effects for TM polarization. In **Fig. 6(b)**, the higher power threshold for TM polarization also leads to a slight increase in the *PER* when ~0 dBm <

$P_{in}$ < ~4 dBm. We also note that the *PER* slightly decreases when $P_{in}$ > ~4 dBm. This is mainly induced by a more dramatic decrease in the *ER* for TM polarization, which has a higher value (in dB) compared to TE polarization. It is also worth mentioning that the results in **Fig. 6** were measured at one of the MRRs' resonance wavelengths near 1550 nm. Similar phenomena were also observed at other resonance wavelengths within the maximum tuning range of our CW laser (*i.e.*, 1500 – 1600 nm).

By using the scattering matrix method[43-51] to fit the measured transmission spectra, we obtained the *PL*'s of the GO-coated segment in the hybrid MRR. The *k*'s of 2D GO films were further extracted from the obtained *PL*'s using the same method as in **Figs. 4(a)**. **Figure 6(c)** shows the *k*'s of 2D GO films versus $P_{in}$, and **Fig. 6(b)** shows the ratios of the TE- to TM-polarized *k* values calculated from **Fig. 6(c)**. For comparison, we also show the corresponding results obtained from the waveguide experiment in **Fig. 4(d)**. In **Fig. 6(c)**, the *k* values obtained from the MRR experiment match closely with those obtained from the waveguide experiment, reflecting the consistency of our GO film fabrication process. As the input power increases, the *k* obtained from the MRR experiment exhibits a more significant increase, indicating that there was more significant photo-thermal reduction of GO in the hybrid MRR due to the resonant enhancement effect. When $P_{in}$ ≤ 0 dBm, the ratio in **Fig. 6(d)** remains nearly constant at ~5.3, which is consistent with that in **Fig. 4(a-iii)**. For $P_{in}$ > 0 dBm, the ratio increases with $P_{in}$, achieving a maximum value of ~6.9 at $P_{in}$ = ~8 dBm. In contrast, the ratio obtained from the hybrid waveguide was still ~5.3 at the same input power.

In summary, waveguide and MRR polarizers are demonstrated by integrating 2D GO films onto Si photonic devices. We fabricate hybrid devices with precise control over the thicknesses and lengths of GO films and perform detailed measurements with them. By optimizing the GO film thicknesses, lengths, and Si waveguide widths, up to ~17-dB *PDL* and ~10-dB *PER* are achieved for the waveguide and MRR polarizers, respectively. The dependence of the polarizers' response on the wavelength and power of input light is also characterized, showing a broad operation bandwidth over ~100 nm as well as polarization selectivity improvement enabled by photo-thermal changes in GO. Finally, we find that the anisotropy in the loss of GO films dominates the polarization selectivity of these polarizers by fitting the experimental results with theoretical simulations. Our study provides an attractive approach for implementing high-performance optical polarizers by integrating 2D GO films onto Si photonic devices and complements the recent advances made in GO coated nanophotonic devices for linear and nonlinear optics [52-74] as well as a wide range of nonlinear optical devices [75-174].

## AUTHOR DECLARATIONS

### Conflict of Interest

The authors have no conflicts of interest to disclose.